\documentclass{ws-mpla}

\begin{document}

\markboth{Hammer et al.}
{Formation of large disks}

\catchline{}{}{}{}{}

\title{THE FORMATION OF LARGE GALACTIC DISKS THROUGH THE HIERARCHICAL SCENARIO: FURTHER CONSEQUENCES\\
}

\author{\footnotesize FRANCOIS HAMMER
}

\address{GEPI, Observatoire de Paris \& CNRS, 5 Place Jules Janssen, 92195 MEUDON FRANCE
\\
francois.hammer@obspm.fr}

\author{YANBIN YANG}
\address{GEPI, Observatoire de Paris \& CNRS 5 Place Jules Janssen, 92195 MEUDON FRANCE}

\author{HECTOR FLORES}
\address{GEPI, Observatoire de Paris \& CNRS, 5 Place Jules Janssen, 92195 MEUDON FRANCE}

\author{MATHIEU PUECH}
\address{GEPI, Observatoire de Paris \& CNRS, 5 Place Jules Janssen, 92195 MEUDON FRANCE}

\maketitle

\pub{Received (10 September 2012)}{Published (11 October 2012)}

\begin{abstract}
Using the deepest and most complete observations of distant galaxies, we investigate the progenitors of present-day large spirals. Observations include spatially-resolved kinematics, detailed morphologies and photometry from UV to mid-IR. Six billions years ago, half of the present-day spirals were starbursts experiencing major mergers, evidence for this is provided by their anomalous kinematics and morphologies. They are consequently modeled using hydrodynamics models of mergers and it perfectly matches with merger rate predictions by state-of-the-art-$\Lambda$CDM semi-empirical models. Furthermore imprints in the halo of local galaxies such as M31 or NGC5907 are likely caused by major merger relics. 
This suggests that the hierarchical scenario has played a major role in shaping the massive galaxies of the Hubble sequence. Linking galaxy properties at different epochs is the best way to fully understand galaxy formation processes and we have tested such a link through generated series of simulations of gas-rich mergers. Mergers have expelled material in galactic haloes and beyond, possibly explaining 60\% of the missing baryons in Milky-Way (MW) mass galaxies. A past major merger in M31 might drastically affect our understanding of Local Group galaxies, including MW dwarfs. We also propose future directions to observationally constrain the necessary ingredients in galaxy simulations.

\end{abstract}

\ccode{PACS Nos.: include PACS Nos.}

\section{Introduction}

The way galaxies form is still not fully understood and no consensus on this question has been reached yet. Galaxies assemble their mass through major, minor mergers, and gas accretion but the balance between these mechanisms is still debated. The present generation of instruments allows us to study with considerable details the progenitors of present-day galaxies up to eight or even ten billion years in the past. Spatially resolved kinematics is currently observed\cite{Flores06,Forster-Schreiber06,Law09,Epinat09} revealing  internal motions on 3 to 7 kpc scales, up to z=2.5. Imagery with the HST-ACS reveals sub-kpc details and numerous instruments allow us to retrieve almost completely their spectral energy distribution from UV to radio wavelengths. Then why a consensus is so hard to establish? \\

Perhaps this is caused by a combination of multiple reasons.  Large scale simulations are remarkable tools to investigate galaxy formation though relating them to observations is not always trivial. Moreover it has been shown\cite{Hopkins10a} that simplifications in semi-analytic models (SAM) may considerably alter their predictions. For example the major merger rate has been underestimated by factors reaching 10 for models assuming (unphysical) instant stripping of the smallest interloper gas. Such an assumption does not only considerably reduce the merger rate experienced by a galaxy, but it also prevents the gas from retaining the orbital angular momentum and then from possibly reforming a disk. A complete review of the various caveats in SAMs can be found in Hopkins et al.\cite{Hopkins10a}. It has lead to significantly improved predictions from cosmological simulations that account for the numerous spiral galaxies observed today\cite{Font11}. 

Observations could also provide contradictory results, often caused by methodological reasons, not the least of them being the differences between the selection criteria. Representative samples of present-day galaxy progenitors at six billion years look-back time\cite{Yang08,Delgado09} (z$\sim$ 0.7) can be securely gathered, but this is still problematic at larger look-back times\footnote{Selecting z$>$1 galaxies using visible filters strongly biases the selection in favour of actively forming, young and low-mass galaxies. Moreover, linking z$>$1 galaxies to their descendants is problematic given the high probability of merger events during the elapsed time, i.e., $³$ 8 billion years.}. Gathering sufficient amount of complementary observations is also vital, because several mechanisms could explain a single observation. For example spatially resolved kinematics should be associated with deep imagery at red rest-frame wavelengths, otherwise without knowing the orientation of the disk major axis, it would be uncertain whether or not a galaxy is rotating. \\
 
Galaxy formation is not only interesting by itself, but is intimately related to the way astrophysicists have established the dark matter (DM) content of the Universe. The spiral galaxy rotation curves and velocity dispersion of MW faint dwarfs provide two important observational lines of evidences on the DM. It may be useful to understand how a galaxy has been formed prior to evaluating its DM content.

In the present paper we will introduce the most accomplished survey of distant galaxies, called Intermediate MAss Galaxies Evolutionary Sequence (IMAGES). In Section 2, we present the morphological and kinematical properties of distant galaxies showing that half of local spiral disks were not in place six billion years ago. In Section 3, we find that this result is consistent with $\Lambda$CDM model predictions, i.e. that the morphologies and kinematics of distant galaxies are consistent with various major merger phases. In Section 4, we verify whether this result is robust with regards to detailed observations of nearby spirals and their haloes. In Section 5,  we investigate some consequences of the hierarchical scenario, within the Local Group and how it contributes to solve the missing baryon problem. Section 6 suggests some directions to improve both observations and simulations, including by bringing IMAGES observational constraints to the models of star formation history  (SFH).

\section{A galaxy evolution census during the past 6 billion years}

The explicit goal of IMAGES  is to gather enough constraints on z=0.4-0.8 galaxies for linking them directly to their descendants, the local galaxies. Its selection is limited by an absolute J-band magnitude ($M_{J}(AB)<$ -20.3), a quantity relatively well linked to the stellar mass\cite{Yang08}, leading to a complete\footnote{The stellar mass distribution of IMAGES galaxies is well consistent with that of the deepest surveys; strictly speaking, IMAGES is the only representative survey of distant galaxies with spatially resolved kinematics because other surveys are either contaminated by ultra star forming dwarves or are not selected using galaxy stellar mass or proxy's.} sample of 63 galaxies with $M_{stellar}>$ 1.5 $10^{10}$ $M_{\odot}$, and with an average value similar to the MW mass. The set of measurements includes:
\begin{itemize}
\item  HST imagery (one to three orbits in b, v, i and z) to recover colour-morphology comparable\cite{Delgado09} to the depth and resolution ($<$400 pc) of local galaxies from the Sloan Digitized Spectroscopic Survey\cite{Abazajian09} (SDSS);
\item spatially-resolved kinematics from VLT/GIRAFFE (from 8 to 24hrs integration times) to sample gas motions at $\sim$ 7 kpc resolution scale\cite{Yang08}; 
\item deep VLT/FORS2 observations (two $\times$ three hours with two grisms at R=1500) to recover the gas metal abundances\cite{Rodrigues08};
\item Spitzer 24$\mu$m observations to estimate the extinction-corrected SF rates as well as Spitzer IRAC and GALEX deep observations providing photometric points to constraint the spectral energy distribution.
\end{itemize}

Taken together, these measurements ensure that the IMAGES sample collects enough data to characterise morphologies, internal motions, and mass evolution of distant galaxies. Their depth and resolution allow us to retrieve essential parameters with a similar accuracy to that which is currently obtained for local galaxies. IMAGES galaxies have been observed in four different fields of view to avoid cosmological variance effects. IMAGES is hardly affected by cosmological dimming, because the imaging depth ensures the detection of the optical disk of a MW-like galaxy after being redshifted to z $\sim$ 0.5. \\

According to the Cosmological Principle, progenitors of present-day giant spirals have properties similar to those of galaxies which emitted their light $\sim$ 6 Gyr ago. 
Fig. 1 presents the results of a morphological analysis of 116 SDSS galaxies (top) and of 143 distant galaxies for which depth, spatial resolution and selection are strictly equivalent\cite{Delgado09}. The 143 distant galaxies (Fig. 1, bottom) include the 63 from IMAGES as well as a complementary number of distant galaxies without emission lines\cite{Delgado09}. Statistical distributions (stellar mass, star formation, etc.) of both low and high redshift galaxies have been verified to be consistent with the results of large surveys. The methodology for classifying the morphologies follows a semi-automatic decision tree, which uses as templates the well known morphologies of local galaxies that populate the Hubble sequence, including the color of their sub-components\cite{Neichel08,Delgado09}. Such a conservative method is the only way for a robust morphological classification, and indeed, the results\cite{Delgado09} are similar to those of an expert\cite{vBergh2002} in the field. 

However morphological appearance may be affected by star formation. The next step in classifying the nature of distant galaxies is to compare the morphological classification to the spatially-resolved kinematics. The latter provides a kinematical classification of velocity fields ranging from rotation, perturbed rotation or complex kinematics\cite{Flores06,Yang08}.  It has been robustly established\cite{Neichel08} that peculiar morphologies correlate well with anomalous kinematics and vice versa: 95\% and 86\% of galaxies with complex kinematics  and perturbed rotations have peculiar morphologies, respectively. On the other hand 80\% of galaxies with robust rotation show spiral morphologies.  

Is that correlation preserved when using automatic classification methods based on concentration and asymmetry of the galaxy luminosity profile? The answer is negative\cite{Neichel08}, and these methods overestimate the number of spirals by a factor of two, a problem already identified\cite{Conselice2005}. Automatic classification methods are interesting and appealing because they can be applied to very large number of galaxies. However they provide too simplistic results to interpret complex objects like distant galaxies. Their limitations in distinguishing peculiar from spiral morphologies lead to a far larger systematic than the Poisson statistical uncertainty for a sample of one hundred of objects\cite{Delgado09}. 

\begin{figure}[h!]
\centerline{\psfig{file=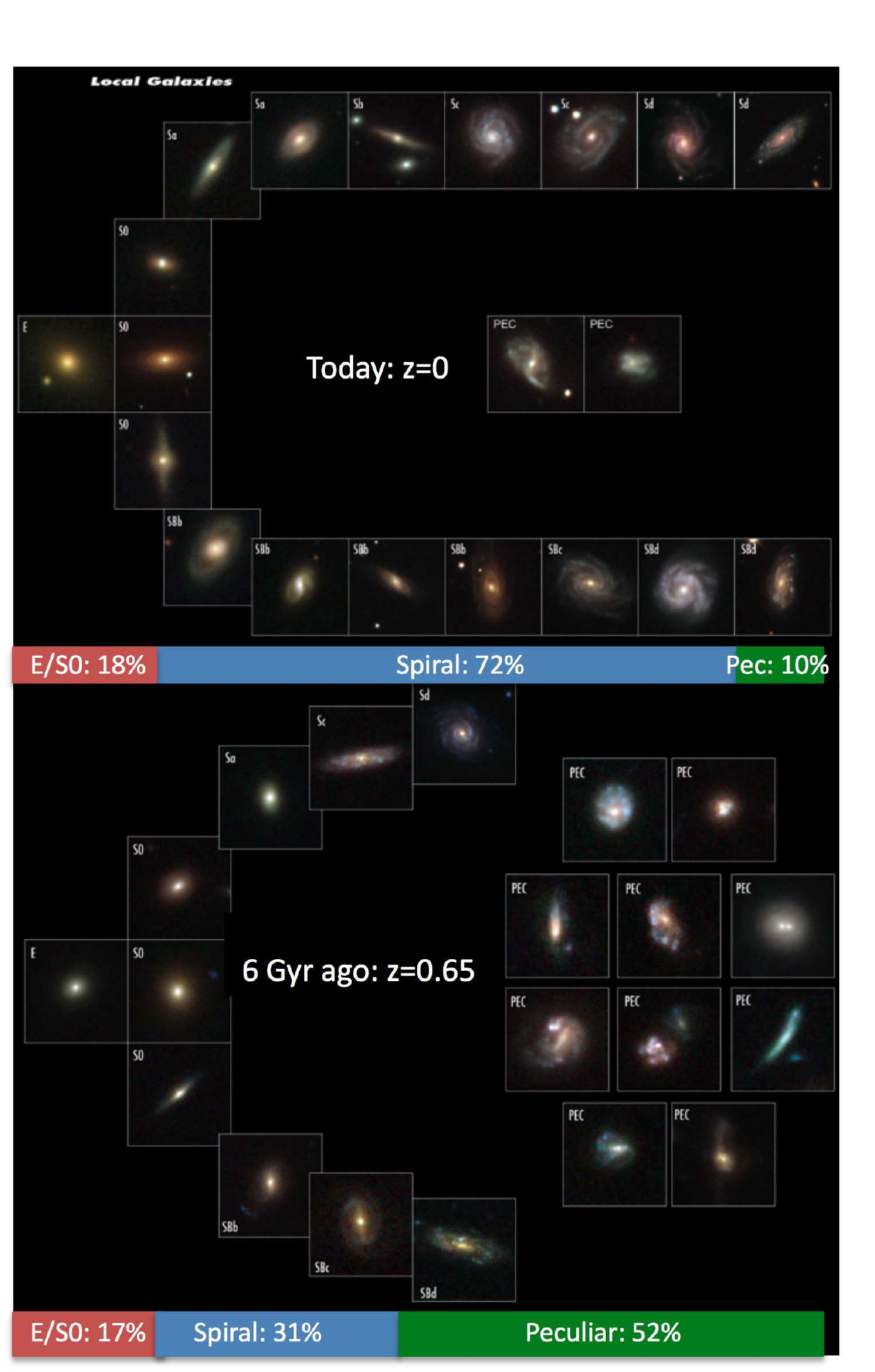,width=4.3in}}
\vspace*{8pt}
\caption{Present-day Hubble sequence derived from the local sample (top) and past Hubble sequence derived from the distant sample (bottom). Each stamp represents approximately 5\% of the galaxy population. Galaxy fractions are given in percentage.\protect\label{fig1}}
\end{figure}

Fig. 1 illustrates the global evolution of the Hubble sequence during the
past 6 Gyr. The link between the two Hubble sequences (past and present-day) is marginally affected by very recent mergers or by stellar population evolution. 
Merger occurrence decreases rapidly at recent epochs and the stellar evolution
is precisely compensated by the redshift evolution of the $M_{stellar}$/$L_{K}$ ratio\cite{Delgado09}. As a result galaxies in the bottom panel of Fig. 1 should be the progenitors of those in the top panel. The fraction of E/S0 galaxies has not evolved, while half of the spirals were not in place 6 Gyr ago, or in
other words, half of the spiral progenitors have either peculiar
morphology and/or anomalous kinematics.   

\section{The observed prominence of mergers in the spiral past history}

The remarkable agreement between morphological and kinematical
classifications implies that dynamical perturbations of the gaseous
component at large scales are linked to the peculiar morphological
distribution of the stars. This suggests a common process at all
scales for gas and stars in these galaxies. Which physical processes
may be responsible of this morpho-kinematic behavior? Most anomalous
galaxies reveal peculiar large-scale gas motions affecting their overall velocity field. Minor mergers (mass ratio well above five) can locally affect the dispersion map\cite{Puech07} while they do not affect the large scale
rotational field over several tens of kpc\footnote{This is entirely true for very minor mergers (mass ratio more than 10) while intermediate mergers may affect significantly the galaxy properties during short phases. Due to their lower impact and their longer duration\cite{Jiang08}, they are considerably
less efficient than major mergers in distorting morphologies and kinematics or they do it in a somewhat sporadic
way\cite{Hopkins08}.}. Outflows provoked by stellar feedback are not observed in the IMAGES galaxies that are generally forming stars at moderate rate\cite{Hammer09b,Rodrigues12}. Internal fragmentation is
limited because less than 20\% of the IMAGES sample show clumpy morphologies\cite{Puech10} while associated cold gas accretion tends
to vanish in massive halos at z$<$1, with rates $<$1.5 $M_{\odot}$/yr at
z$\sim$ 0.7\cite{Keres09,Brooks09}. Finally perturbations from
secular and internal processes (e.g. bars or spirals) are too small to be detected by the "large-scale" spatially resolved spectroscopy of IMAGES. Major mergers appear to be the most likely mechanism explaining abnormal morphologies as well as peculiar large scale motions of the gas in anomalous galaxies. In fact merger random walk evolution explains the strong redshift increase in the scatter of both the mass-velocity (Tully-Fisher) relation\cite{Flores06,Puech08,Covington10} and the luminosity-metallicity relation\cite{Liang06,Rodrigues08}.

This has led us to test, and then successfully model, five of the
IMAGES galaxies via major mergers\cite{Peirani09,Yang09,Hammer09a,Puech09,Fuentes10} using
hydrodynamical simulations (GADGET2 and ZENO). However the amount of
data to be reproduced per galaxy is simply enormous, leading to 21
observational constraints to be compared to 16 free model parameters\cite{Yang09}. We have limited our
subsequent analysis to the 33 anomalous galaxies belonging to one field for
reasons of data homogeneity. A comparison of their morpho-kinematics
properties  to those from a grid of simple major merger models\cite{Barnes02}, provided convincing matches in about two-thirds of
the cases\cite{Hammer09b}. This implies that a third of z=0.4-0.8 spiral galaxies are
(or have been) potentially involved in a major merger. Setting aside the bulge-dominated galaxies (E/S0) which fraction does not evolve, means that half of present-day spirals were involved in a major merger phase six billion years ago\cite{Hammer09b}. \\

Why IMAGES is finding so many galaxies that can be attributed to a major merger phase? In fact the morpho-kinematic technique used in IMAGES is found to be sensitive to all merger phases, from pairs to the post-merger relaxation phase.  The merger rate associated with different phases has been calculated\cite{Puech11}, and found to match perfectly predictions by state-of-the-art $\Lambda$CDM semi-empirical models \cite{Hopkins10b} with no particular fine-tuning. Thus, both theory and observations predict an important impact of major mergers for progenitors of present-day spiral galaxies so that the whole Hubble sequence could be just a vestige of merger events\cite{Hammer09b}.

\section{Galactic bulges and disks as relics of the last major merger}

The present-day Hubble sequence (see Fig.1 top panel) indicates a remarkable behavior of massive galaxy morphologies. Almost all are structured along two main sub-components, a dispersion dominated bulge surrounded by a rotationally supported disk. Why are such complex systems so well organised along a decreasing bulge-to-total ratio (B/T) if they are affected by numerous internal and external physical processes? This suggests a single dominant process\cite{Hammer09b}, and mergers could be the culprits as the last major event roughly traces B/T\cite{Hopkins10b}. 
However because major mergers can easily destroy thin rotating disks\cite{Toth92}, this creates an apparent tension between the large fraction of present-day disks and their survival within the $\Lambda$CDM\cite{Stewart09}. Solving this requires that many disks should be rebuilt after mergers, which is expected if the initial gas content is large enough. If true, mergers may bring enough orbital angular momentum to solve\cite{Maller02} the spin catastrophe, i.e. the longstanding problem that disks formed in isolation are too small and with too small angular momentum\cite{Navarro00}. 

 Athanassoula\cite{Athanassoula10} reviewed the different physical processes leading to the formation of elliptical and spiral galaxies. The "merger hypothesis"\cite{Toomre72} for elliptical galaxies interprets them as the product of gas-poor or successive major mergers. It appears more and more plausible that spiral galaxies could also result from mergers of gas-rich galaxies\cite{Hammer05,Hammer09b}. Re-formation of disks after major mergers has already been described in 2002\cite{Barnes02}, assuming a small gas fraction (12\%) in the progenitors. With much larger gas fraction a rebuilt disk could be more prominent and after subsequent star formation, it could dominate the galaxy\cite{Brook04,Springel05,Robertson06} as shown in Fig. 2. In fact estimates of gas fractions in galaxies at earliest epochs indicate values that exceed 50\% at z$\sim$ 1.5-2\cite{Daddi10,Erb06,Rodrigues12}. These are likely low limits because observations of very distant galaxies are biased towards either massive or star-forming objects.


\begin{figure}[h!]
\centerline{\psfig{file=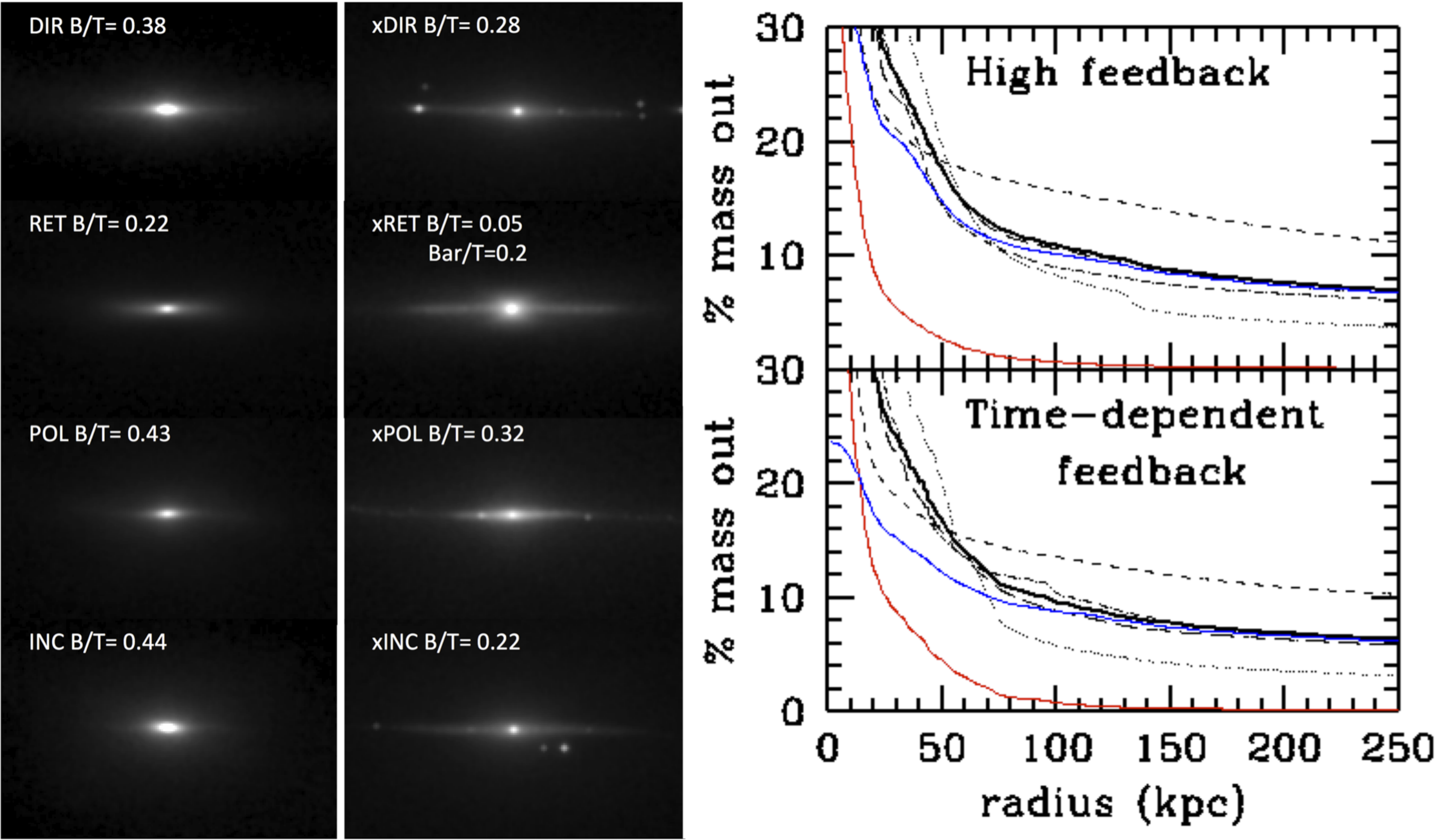,width=5.0in}}
\vspace*{8pt}
\caption{{\it Left:} Simulated 3:1, gas-rich (initial gas fraction: 60\%) merger remnants, 8 billion years after fusion, assuming Barnes (2002, see Table 1 of Ref. 30)
 orbits with pericentre radius of 16kpc. Each image is 35 kpc wide, horizontally. The model is based on GADGET2 within which we have implemented cooling, star formation, and feedback processes following Cox et al. (2006) $^{\rm 45}$.
However we have adopted a large number of particles, i.e. 1500K particles including 500K baryons with a corresponding mass similar to that of the MW (7.3 $10^{10}M_{\odot}$) and a baryonic fraction of 20\%. DM and baryon particles have masses of 2.75 $10^{5} M_{\odot}$ and 1.375 $10^{5} M_{\odot}$, respectively.  {\it Right:} Fraction of baryons outside a given radius for merger remnants 8 billion years after fusion. The thick black line represents the average of DIR (dots), RET (short dash), POL (long dash), and INC (dot-short dash) orbits. Blue and red solid lines represent the averaged, outside fraction of gas and stars to total mass.\protect\label{fig2}}
\end{figure}

 Empirical models of galaxy luminosity or stellar mass function and gas content allow\cite{Hopkins10b} us to evaluate which mergers matter. The hierarchical scenario reproduces well the observed increase of B/T with luminosity or mass\cite{Hopkins10b}. The derived merger rate implies that major mergers could dominate the formation and assembly of MW mass galaxy substructures\cite{Hopkins10b}. The corresponding major merger rate\cite{Hopkins10b} with baryonic mass ratio smaller than 4:1, provides values consistent with IMAGES observations\cite{Puech11}. It also predicts that a MW mass galaxy (as those in the top panel of Fig. 1) should have experienced on average 1.2  such events during the past 12.5 billion years (z=6). Then it is possible that the overall morphology of massive galaxies has been shaped by major mergers: in fact an increasing number of recent cosmological simulations indeed lead to the formation of late-type disk galaxies with MW mass after such events\cite{Font11,Brook11}.\\

A major merger scenario may face the problem of reconstructing disks as thin as the observed ones. In fact, numerical simulations are rapidly progressing, and the recent  simulations\cite{Keres11,Guedes11,Hopkins12} show rebuilt thin disks after mergers that could resemble the observed ones.  On the observational side, near-IR observations of edge-on spirals\cite{Comeron11} have revealed the presence of massive thick disks. This could relax the thin disk requirement from observations because near-IR morphology provides the best proxy of the stellar mass distribution\cite{Wang12}. Non negligible thick disks are indeed expected to form after major mergers\cite{Hammer10,Wang12}.

Perhaps the most difficult challenge for the hierarchical scenario is due to the presence of very late type spirals, i.e. bulge-less galaxies. It has been shown\cite{Kormendy10} that within 8 Mpc, most massive galaxies are bulge-less or have bulges that follow a luminosity profile similar to that of the disk\footnote{Such bulges are called pseudo-bulges or disky-bulges as opposed to classical bulges that show a very centrally concentrated mass distribution and the formation of which is associated with gas-poor or successive mergers.}. Part of this problem may be alleviated because the B/T ratio strongly depends on the merger orbital parameters. Fig. 2 presents rebuilt disks 8 billion years after a gas-rich merger and the B/T ratio may vary by a factor 2 from prograde-prograde (DIR) to retrograde-retrograde (RET) orbits. Adopting low feedback values\cite{Cox06} after the fusion (models labelled "x") also helps to reconstruct more dominant disks. Alternatively the feedback may be assumed to be more efficient within the central region\cite{Governato09,Brook11} allowing for redistribute of the gas angular momentum to the disk. Many of the simulated bulges displayed in Fig. 2 are pseudo-bulges\footnote{Models displayed in Fig. 2 have adopted very low softening lengths (from 10 to 20 pc) allowing to fully sample the central bulge luminosity profile.}, confirming the result\cite{Brook11,Keselman12} that gas-rich major mergers may lead to purely disky-structures, which could resolve the above challenge.




\section{Merger imprints in observations of nearby spirals}
Having a tumultuous merger history 6 Gyr ago should have left some imprints in many present-day spiral galaxies. Let us consider our nearest neighbour, M31 whose classical bulge and high halo metallicity both strongly suggest a major merger origin\cite{vandenBergh05}.  In fact the considerable number of streams in the M31 haunted halo could be the result of a major merger instead of a considerable number of minor mergers\cite{Hammer10}. This provides a robust explanation of the stellar Giant Stream\cite{Ibata01}: 
it could be made of tidal tail stars captured by the galaxy gravitational potential after the fusion time.  In fact stars of the Giant Stream\cite{brown07} have ages older than 5.5 Gyr, which is difficult to reconcile with a recent minor merger\cite{Font08}. A 3:1 gas-rich merger may reproduce\cite{Hammer10} the M31 substructures (disk, bulge \& thick disk) as well as the Giant Stream assuming the interaction and fusion may have occurred 8.75$\pm$0.35 and 5.5 $\pm$0.5 Gyr ago, respectively. Besides this, the Milky Way may have had an exceptionally quiet merger history\cite{Hammer07} during the past 11 billion years.

Further away from the Milky Way, the outskirts of isolated spiral galaxies have been observed\cite{Martinez10} at unprecedented depth, up to a surface
brightness sensitivity of $\mu_{V}$= 28.5 mag/arcsec$^2$. This work revealed stellar streams of various shapes surrounding many spiral galaxies. These observations are often considered as evidencing the presence of minor mergers in spiral galaxies. For example, the NGC5907 halo reveals spectacular and gigantic loops, which have been modeled by a very minor merger  with a 4000:1 mass ratio\cite{Martinez08}. However to reproduce them, minor mergers with mass ratio larger than 12:1 require initial orbits with low eccentricities\cite{Martinez08}, i.e. not consistent with cosmological simulations\cite{Wang12}. Alternatively the whole NGC5907 galaxy (disk, bulge and thick disk) and associated loops have been successfully modeled\cite{Wang12} by assuming a 3:1 gas-rich major merger 8 to 9 billion years ago, and considering orbits consistent with cosmological simulations.  

Further work is needed to firmly establish which process is responsible for the tumultuous history of nearby spirals that is imprinted into their haloes. In most cases\cite{Martinez10}, there is no hint of the residual of a satellite core that would be responsible of the nearby spiral halo streams. Because these structures have sizes comparable to, or larger than, the optical disk of the host galaxy, some satellite cores should be caught during the action, otherwise this favours a major merger origin. 

\section{Further consequences to explore}

\subsection{The nature of MW dwarfs}

A major merger, that occurred five to six billion years ago at the M31 location, may have affected the whole content of the Local Group, because two-third of its baryons lie in this galaxy. In fact gas-rich major mergers are efficient in forming tidal dwarfs\cite{Fouquet12} (TDGs) with baryonic masses up to that of the Large Magellanic Cloud. The proximity of the Magellanic Clouds (MCs) near the MW is quite exceptional\cite{Robotham12}: only 0.4\% of local galaxies display such an environment, and those are found in a double galaxy system such as the Local Group with M31 and the MW. Tracing back the LMC provides several configurations for which it was close to M31 5 to 7 billion years ago\cite{Yang10}. The other exceptional feature is the distribution of MW dwarfs that lie, and have their motions, within a thick plane almost perpendicular to the MW disk\cite{Metz08}. Linking the M31 merger model to the MW leads to a natural explanation of both MCs proximity and the plane of MW dwarfs\cite{Fouquet12}: these would have been formed within a gas-rich tidal tail formed at the first passage of the M31 merger event. Such a tidal tail is inserted within a thick plane that is almost perpendicular to the orbital angular momentum, as is for the M31 disk\cite{Hammer10,Wang12}. Because the later is almost seen edge-on, the thick plane includes the MW. Moreover, the location of the Giant Stream further limits the volume swept by the tidal tail, that still includes the MW. The MW could be presently reached  by the tidal tail, whose modeling predicts a velocity consistent with that of the LMC\cite{Fouquet12}.


The above scenario could appear speculative and is far from being proved. First, MW dwarfs, with the noticeable exception of MCs, are gas-free objects while TDGs expelled by the M31 merger are 90\% gas-rich\cite{Fouquet12}. Second  it has been argued that, due to their large velocity dispersions, MW dwarves are largely dominated by dark matter\cite{Walker08,Strigari08}, while TDGs are DM free.  The first problem has been addressed\cite{Mayer07,Mayer10} through simulations of gas-rich, dwarf irregular galaxies entering the MW potential of a massive galaxy: they can be entirely stripped of their gas through tidal shocks and ram pressure. However such an attempt requires to be done for DM free TDGs, instead of DM-dominated irregular galaxies. Because only the DM-content of  MW dwarfs has been tested, the second problem is of tremendous importance\cite{Kroupa12} for the $\Lambda$CDM scenario that predicts a very large number of DM dominated dwarfs. Simulations\cite{Kroupa97,Casas12} of DM free satellites have shown that, after several orbits around the MW, satellites are out-of-equilibrium bodies with high apparent mass-to-light ratios. If such progenitors of MW satellites have reached the MW potential at  the same time along a few specific orbits, they may reproduce most of their intrinsic kinematics and surface brightness\cite{Kroupa12}. Our simulations of M31 tidal tails\cite{Fouquet12,Yang12} provide realistic TDGs that are far less compact than assumed in the previous work\cite{Kroupa97,Casas12} and are more easily disrupted after a single passage. Completing such a study\cite{Yang12} is mandatory to verify whether or not the large mass-to-light ratios of the MW dwarfs falsify the M31 scenario.

\subsection{Which fraction of missing baryons are expelled during mergers?}

It has been realised that half of the baryons are missing\cite{Fukugita98} in the Local Universe when compared to their amount in the Lyman $\alpha$
forest, at z$\sim$ 3. MW-mass galaxies are also missing 50\% of their baryonic content when compared to the 17\% universal baryon fraction\cite{Sommer-Larsen06,McGaugh10}. Where are the missing baryons? Inventory of baryons in the MW and nearby spirals has
recently been reviewed\cite{Putman12} accounting for the most recent results in all halo gas phases. Neutral hydrogen in high velocity clouds within 50 kpc provides less than $10^{8}M_{\odot}$   i.e., less than a percent of the disk baryonic mass.  Lyman $\alpha$ absorbers show a high covering fraction of  warm and hot hydrogen gas with $\sim$ $10^{10}M_{\odot}$ within 300 kpc\cite{Prochaska11}, i.e., 20\% of the MW-disk baryonic mass.  Detection of gas in very hot phases requires X-ray spectroscopy and is often limited by the difficulty to estimate the distance from the MW center. Perhaps, from a recent detection within a 200 kpc radius\cite{Gupta12}, it could reach the baryonic mass of the MW disk. Thus the missing baryons are likely to be located beyond 50 kpc and deeper observations at all wavelengths are necessary to detect them all. \\

Fig. 2 (right) provides the fraction of baryonic mass ejected by major merger models. It appears that 18\% of the total baryonic mass is distributed outside a radius of 50 kpc while 7\% lies outside 250 kpc, a radius often adopted for the virial radius of the MW. In Fig. 2 the baryonic mass of the outer halo and beyond is essentially made of gas, and its fraction does not vary significantly with the orbits or with the star formation history. 
Because each MW-like galaxy has experienced 1.2 major merger over the past 12.5 billion years, possibly $\sim$ 22\% of the baryons lie beyond 50 kpc, confirming previous results about the missing baryons driven by major mergers\cite{Sinha09,Sinha10}. This fraction of baryons in galaxy's outskirts can be entirely attributed to mergers because we have assumed almost no baryons beyond 50 kpc in the initial interlopers. Minor mergers may also contribute to the expulsion of baryons: a 7:1 minor merger expels  8.5 and 4\% of the baryons at 50 and 250 kpc, respectively. Integrating the merger rate to $\le$ 10:1 mergers implies that up to 30\% of the whole baryonic mass of MW-mass galaxies could have been ejected at more than 50 kpc.\\

These preliminary results need to be extended to a wider galactic mass range. Such an estimate\cite{Sinha10} has been previously attempted, although it only accounted for the few percent of unbound hot baryons expelled beyond the virial radius. We confirm this result but find a much larger value if accounting for cold and warm gas expelled through tidal tails (10\% at $>$ 250 kpc). In fact the baryon budget\cite{Fukugita98} only compares the detected to the expected baryonic contents, and most missing baryons likely lie within the virial radius\cite{Sommer-Larsen06}, as also predicted by Fig.2. Our simulations show that the fraction of expelled baryons beyond 50 kpc is very stable with time, and mergers could be responsible of $\sim$ 60\% (corresponding to 30\% of the whole baryonic content) of the missing baryons in MW-mass galaxies. Accounting also for the history of gas accretion, feedback \& heating from a cosmic UV field\cite{Keres11,Brook11} could be sufficient to explain the all missing baryons.




\section{Prospective and conclusion}

Important progress is expected on both observational and theoretical sides: IMAGES results have to be independently confirmed, although it needs to avoid automatic procedures that often degrade the significance of astrophysical data. With this, the so numerous unstable, anomalous progenitors of present-day spirals, with behavior so similar to major mergers will provide major constraints for disk galaxy formation theories and simulations.\\

\begin{figure}[h!]
\centerline{\psfig{file=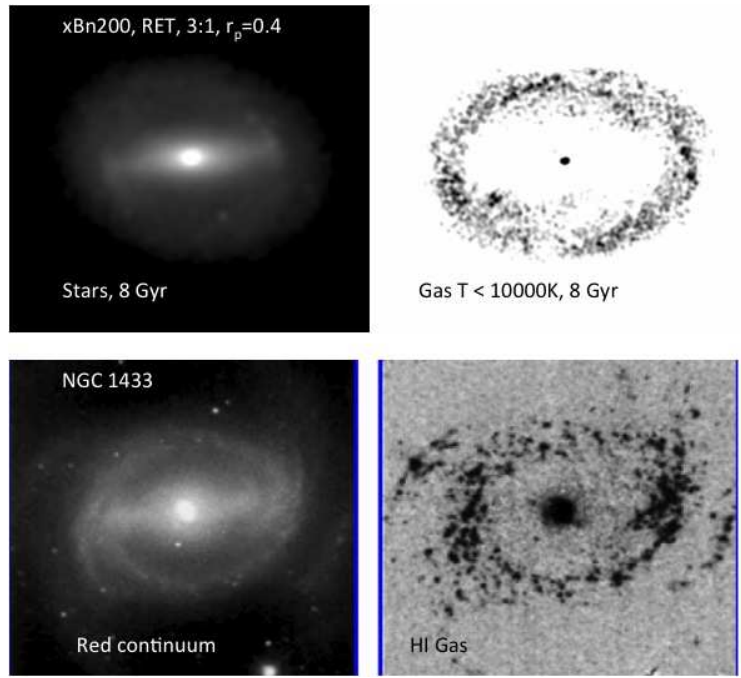,width=4.0in}}
\vspace*{8pt}
\caption{{\it Top}: Central view of the remnant of a retrograde (RET$^{\rm 30}$) orbit (almost face-on view of Fig. 2, Left panel, xRET) for stars and cold gas. The formation of the bar follows quite accurately the star formation in the rebuilt disk, and it appears 1.5 billion year after the fusion. {\it Bottom}: Observed red stars and HI gas of NGC 1433$^{\rm 83}$.\protect\label{fig3}}
\end{figure}

Further investigations of the merger role in shaping galactic substructures are necessary. Fig. 3 shows an example of a simulation adopting a very small gravitational softening length, $\epsilon$= 10 pc, allowing the investigation of central regions of the remnant that show some similarities with the nearest classic ringed barred spiral, NGC 1433.  About one fifth of all spirals include a ring or pseudo ring-shaped pattern in the light distribution, often associated with a bar\cite{Buta96}. Ultimately using series of major merger simulations it may be possible to reproduce the current optical and HI morphologies of each spiral in the Hubble sequence. Fig. 3 illustrates the need for very large number of particles to resolve both central and outer parts. This could favour our approach that consists to model individual merger allowing to detail small scale structures (e.g., bulge \& bars) rather than using zoomed cosmological simulations for including smooth gas accretion from the intergalactic medium. The latter will be systematically less efficient to sample small scales because they would need extremely high number of particles\cite{Brook11}. 

Our modeling should be up-graded to include gas accretion, perhaps in following simple fitting formulae for the cold gas accretion rate\cite{Faucher12}. This is why we have limited our simulations to 8 billion year after fusion, i.e., a period (z$\le$ 1.07) during which cold streams penetrating the hot atmospheres of massive halos at z$\ge$ 2 gradually disappear\cite{Keres11}. Such a period is sufficient to investigate a significant part of the mergers having lead to spiral morphologies. 

Implementing the feedback history is a well-known problem for galaxy simulations. Fig. 2 shows two-set of models with a feedback efficiency half of the high feedback defined by Cox (2006)\cite{Cox06}, before the fusion, one set assuming this feedback for all the simulation duration, the other being time-dependent, assuming a decrease after fusion to a low\cite{Cox06} feedback value. This may be crude even if it has been quite compelling for reproducing the details of galaxy substructures and halo streams in M31 and NGC 5907. Considerable progress has been made in evaluating the feedback and its role in the gas-to-stars transition\cite{Hopkins12}. Large feedback from early massive stars and observations at high z qualitatively support an early large feedback efficiency. 
IMAGES includes sufficiently large number of mergers at different phases to sample a star formation history mimicking that associated to mergers\cite{Puech11}. This crucial observation could be used  to empirically calibrate the feedback efficiency, assuming that feedback is controlling the SFH\cite{Hopkins12}.\\

Why is it so important to link distant to nearby galaxies? The present stage of identifying galaxy morphological classes (e.g., Fig. 1) and sub-classes (e.g., Fig.3) is not sufficient to explore the galaxy formation past history. The proposed link could improve our understanding of galaxy formation with an impact comparable to the discovery of the DNA for interpreting the  zoological classification. Moreover, during mergers, a significant fraction of cold baryons are ejected in the halo, retaining some orbital angular momentum, and in fact HI gas often displays large-scale spiraling structures. Perhaps spiral galaxies are not in equilibrium, thus affecting the DM estimates from their HI rotation curves. Besides this, Fig.2 illustrates that different SFHs lead to very different number of TDGs produced by a merger event: this requires one to constrain the SFH prior to an attempt to evaluate the fraction of dwarfs that could be TDGs\cite{Okazaki00}. Further observations from currently existing telescopes are required, including:
\begin{itemize}
\item  A complete sample of z$>$2 galaxies selected from K band for sampling the red stellar population at rest. This is the only way to quantitatively estimate the relative roles of mergers and gas accretion at this epoch, and this requires an observational data set similar to that provided by IMAGES.
\item To derive the unknown fraction of low surface brightness galaxies at z= 1-2. Those are rare objects locally, but their number fraction may have increased dramatically in the distant, gas-dominated Universe. This is crucial for testing the merger progenitor properties that are otherwise assumed to be thin disks.
\item The best place to investigate relics deposited by mergers and other phenomena is the halo: ionised gas properties in galactic haloes can be derived from integral field units up to z$\sim$ 0.7, and HI extended maps are currently available\cite{Walter08} in a 15 Mpc local volume.
\end{itemize}

\section*{Acknowledgements}
We are very indebted to Sidney van den Bergh for his thorough reading of the paper and for providing us with a far improved English wording.

\end{document}